# Experimental realization of topological on-chip acoustic tweezers

Hongqing Dai,[1, 3, a)] Linbo Liu,[2, 4, a)] Baizhan Xia[1, b)] and Dejie Yu[1, b)]

1. State Key Laboratory of Advanced Design and Manufacturing for Vehicle Body, Hunan University, Changsha, Hunan, 410082, People's Republic of China
2. School of Mechanical Engineering, and Jiangsu Key Laboratory for Design and Manufacture of Micro-Nano Biomedical Instruments, Southeast University, Nanjing, Jiangsu, 211189, People's Republic of China
3. Biomaterials Innovation Research Center, Division of Biomedical Engineering, Brigham Women's Hospital, Harvard Medical School, Cambridge, Massachusetts,02139, United States
4. John A. Paulson School of Engineering and Applied Sciences, Harvard University, Cambridge, Massachusetts, 02138, United States
a) Contributions: Hongqing Dai and Linbo Liu contributed equally to this work.
b) Author to whom correspondence should be addressed: xiabz2013@hnu.edu.cn, djyu@hnu.edu.cn

**Abstract**

Acoustic tweezers are gaining increasing attention due to their excellent biological compatibility. Recently, the concept of topology has been expanded from condensed matter physics into acoustics, giving rise to a robust wave manipulation against defects and sharp turns. So far, topological acoustics have not been experimentally realized in on-chip level which can be worked as tweezers for microparticle manipulations. Here, we achieved a topological on-chip acoustic tweezer based on the topologically protected phononic mode. This tweezer consisted of one-dimensional arrays of Helmholtz resonant air cavities. Strong microfluidic oscillations induced by acoustic waves were experimentally observed at water-air surfaces of Helmholtz resonant air cavities at the topological interface. Acoustic radiation force induced by these microfluidic oscillations captured microparticles whose sizes were up to 20 µm and made them do orbital rotations. Our topological on-chip acoustic tweezer realized non-contact label-free microparticle manipulations in microfluidics and exhibited enormous application potential in the biomedical field.

**Keywords: Topological insulators, Microfluidics, Particles manipulation**

**Introduction**

Acoustic tweezers are versatile tools using sound waves to manipulate particles ranging from nanometer-scale to millimeter-scale in fluidics (1, 2). Due to their non-invasive, label-free, low-power, and biocompatible natures, acoustic tweezers have shown excellent application prospects in cell separation (3), exosome isolation (4), anisotropic muscle tissue culture (5) and acoustically powered microrobots (6). The essential characteristics of acoustic tweezers are their acoustic radiation forces (7, 8) and acoustic streaming forces (9, 10) which drive microparticles to do various motions, from separating, patterning, concentrating to rotating. There are three crucial sorts of acoustic tweezers. Both standing-wave tweezers (7) and traveling-wave tweezers (8) are generally used to realize the translational operation of cells limited by acoustic radiation forces, whereas acoustic streaming tweezers can perform rotational operations via fluid flows generated by oscillating microbubbles (9, 10). The microbubbles seriously suffer from machining errors of devices, perturbations of initial states and aggressiveness of the microfluidic environment. Thus, the vibration characteristics of oscillating microbubbles are unstable, leading to limited reproducibility of acoustic streaming tweezers for fluid mixing and pumping, 3D rotation of cells and small organisms, and neural simulation (10, 11). One promising technique giving rise to a robust wave manipulation is the topological state, which is discovered in condensed matter and recently expanded to acoustics. Essentially different from trivial eigenstates, the topologically protected edge states have a strong resistance to manufacturing defects, environmental disturbances, and even irregular waveguides with blocked cavities and sharp turns. This is very helpful for suppressing the deterioration of required vibration properties and achieving the high-efficient microparticle manipulation under complex, varying and aggressive working conditions.

Topologically protected edge states are generated at the interface between spatial domains with distinct topological phases. There have been considerable efforts to realize such states in acoustics, due to their potential applications in robust acoustic waveguide(12-14), delay lines(15), negative refraction(16) and directional antennas(17). Topological edge states of acoustics were firstly observed in dynamic acoustic networks with circulating fluid to break the time-reversal symmetry(18, 19), and then expanded to time-reversal acoustic systems by simulating valley Hall effects(12, 20) and pseudospin Hall effects(13). Up to now, topological acoustics are only experimentally realized in macroscopic systems, as they can be easily fabricated and tested. On-chip topological insulators, which have achieved tremendous progresses in micromechanical, nanoelectromechanical and nanophotonic systems(21-25), have still not been developed in acoustic systems. The experimental realization of on-chip topological acoustics in microfluidics will provide a new platform for the robust control of acoustic waves in microscale, working as tweezers for microparticle manipulations which have shown significant potential in biochemical reaction, targeted drug delivering and multicellular tissue establishment.

We have designed topological on-chip acoustic tweezers which consisted of Helmholtz resonant air cavities arraying as Su–Schrieffer–Heeger models (26, 27). Due to the surface tension, narrow branch channels of cavities were closed, presenting as a series of connected periodic air chambers (video 1, *supporting information*). Topological edge states of tweezers were generated in interfaces between expanded and shrunk Helmholtz resonant air cavities. We experimentally observed the strong vibrations of water-air surfaces of Helmholtz resonant air cavities which arrayed along topological interfaces, as the acoustic energy was well concentrated in these cavities. Acoustic radiation forces and acoustic streaming forces yielded by vibrating water-air surfaces were used to manipulate microparticle motions. When sizes of microparticles were less than the threshold value which was 25.8 μm in our tweezers, acoustic streaming forces dominated microparticle motions. In this case, microparticles made orbit rotations near water-air surfaces, which was experimentally verified by microparticles with 20 μm. Differing from microbubble-based tweezers, our topological on-chip acoustic tweezers exhibited great robustness against manufacturing defects and uncontrolled environmental disturbances. Furthermore, Helmholtz resonant air cavities were operated in a subwavelength scale, leading that our tweezes can manipulate much larger microparticles when compared with conventional tweezers. Therefore, our topological on-chip acoustic tweezers provided robust and subwavelength techniques for microparticle manipulations.

**Geometric structures**

As shown in Figure 1(a), we plot the schematic of the topological on-chip acoustic tweezer. A piezoelectric transducer, attached to a glass slide adjacent to the polydimethylsiloxane (PDMS) channel, generates acoustic waves, playing a role as a driving source. When the Helmholtz resonant air cavities arranging the inside of the microfluidic channel are exposed to an acoustic field with a wavelength much larger than their geometrical sizes, these cavities at the topological interface will oscillate, especially at the eigenfrequency which is topologically protected. The polystyrene microparticles with diameters of 20 µm are injected into the microfluidics channel. These microparticles are attracted by the oscillation air cavities and carry out orbit rotations. In Figure 1(b), we present the experiment configuration of the microfluidics channel. Along the *y*-direction, the microfluidics channel mainly contains three parts, main channel, square cavities, and narrow channel. The middle square cavities are connected with main channel and narrow channel through the upper and lower necks, respectively. The detailed experimental platform and geometrical sizes of the microfluidics channel are given in Figure S1 (Note 1, *supporting information*). We take two Helmholtz resonant air cavities as a meta-molecule, as shown by the red and blue dash rectangles. The length of the rectangle is $2a = 2$ mm. Along the *x*-direction, the channel can be divided into two parts. The left-part is shrunk, and the distance between two Helmholtz resonant air cavities is $0.8a$. The right-part is expanded with a distance $1.2a$. The upper main channel is filled with water, while the lower narrow channel and middle cavities are filled with air. Each Helmholtz resonant air cavity can be analogous to an inductor-capacitor circuit, with the enclosed cavity acting as the capacitor with capacitance $C \propto V / \rho c^2$, and the neck acting as the inductor $L \propto \rho(L_{eq} / A)$, where $V$ is the volume of the cavity, $\rho$ is the density of the air, $c$ is the sound speed in the air, $L_{eq}$ is the effective length of the neck, and $A$ is the cross-sectional area of the neck. The resonant frequency of a single Helmholtz resonant air cavity is $f = \dfrac{1}{2\pi\sqrt{LC}} \propto \dfrac{c}{2\pi}\sqrt{\dfrac{A}{VL_{eq}}}$. The waterflooding process is shown in Figure 1(c) and videos 1 and 2 in *supporting information*. We slowly inject water into the main channel via a syringe pump. Water fills the whole channel from left to right. Due to the surface tension, water is bounded at the upper narrow necks forming water-air surfaces, instead of flowing to the square cavities.

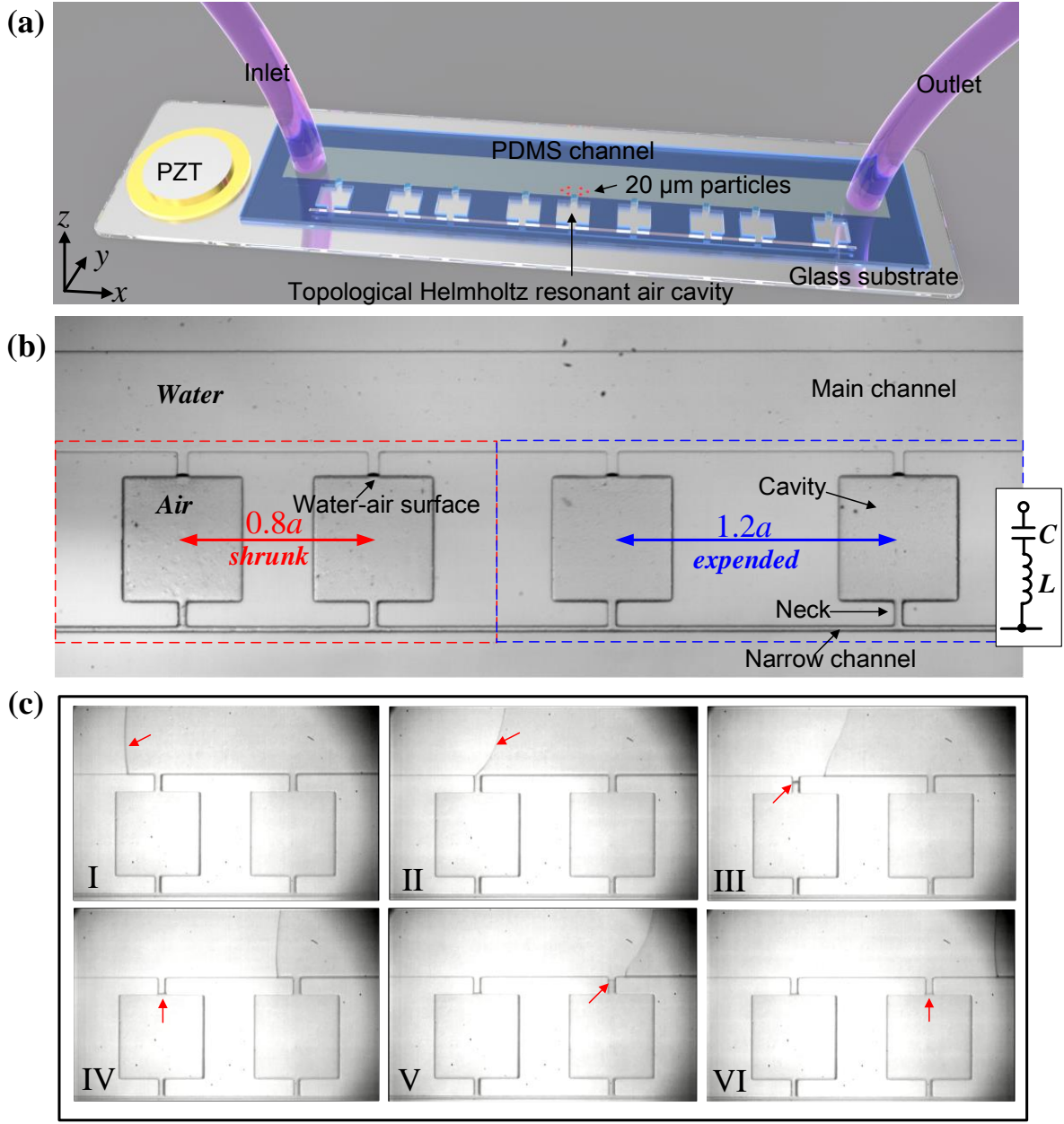

Figure 1. (a) Schematic of the topological insulators in microfluidics chip. (b) Experiment configuration of the microfluidics channel. The main channel is filled with water, and the Helmholtz resonant air cavities are filled with air. The left meta-molecule is shrunk, and the right meta-molecule is expanded. (c) The waterflooding process. The red arrows represent the water-air surface.

**Band structures and topological phase transition**

Considering the large difference of acoustic impedance ($Z = \rho c$) among water (1000 kg/m$^3$, 1480 m/s), PDMS (965 kg/m$^3$, 1050 m/s) and air (1.2 kg/m$^3$, 340 m/s), we can only consider the acoustic field in the air medium with a high reflection induced by water and PDMS. We plot acoustic metamaterials consisted of Helmholtz resonant air cavities, as shown in Figure 2(a). The black dot rectangle represents the unit cell with a Helmholtz resonant air cavity, and the lattice constant is $a$ ($a$ = 1 mm). The red dot rectangle represents the meta-molecule with two Helmholtz resonant air cavities, and the lattice constant is $2a$. In this paper, we calculate the band structures of metamaterials using the band theory in conjunction with the Floquet boundary condition via the finite element method (Simulation, *supporting information*). We firstly calculate the band structure of the unit cell with a Helmholtz resonant air cavity. The first Brillouin zone is [ $-\pi/a$, $\pi/a$ ], and the first band is plotted in Figure 2(b). In Figure 2(c), we calculate the band structure of a meta-molecule with two Helmholtz resonant air cavities. The first Brillouin zone of the meta-molecule is [ $-\pi/2a$, $\pi/2a$ ] which is folded from the first Brillouin zone of the unit cell with [ $-\pi/a$, $\pi/a$ ]. In detail, due to the translational symmetry of metamaterial, the band (marked by left blue curve) in the Brillouin zone [ $-\pi/a$, $-\pi/2a$ ] of the unit cell is translated to the Brillouin zone [ 0, $\pi/2a$ ] of the meta-molecule, and the band (marked by right blue curve) in Brillouin zone [ $\pi/2a$, $\pi/a$ ] of the unit cell is translated to the Brillouin zone [ $-\pi/2a$, 0 ] of the meta-molecule. Therefore, we can get the new folded band structure with a double degenerated point at around 19.74 kHz, as shown in Figure 2(c). The double degenerated point consists of two modes, the monopolar state and the dipolar state, respectively. When the meta-molecule reconstructed from a perfect one with $d=a$ to a shrunk one with $d=0.8a$ (Figure S2, *supporting information*), the degenerated point will be lifted. In Figure 2(d), we calculate the band structure for the meta-molecule with $d=0.8a$. We can find that there is a bandgap from 18.3 kHz to 21.7 kHz. Conventionally, the frequency of Bragg scattering band gap of the ordinary phononic crystal without Helmholtz resonant air cavity is around $f = c/2a$ ($f \approx 171.5$ kHz). However, the wavelength of the band gap obtained in Figure 2(d) is about 1/8 of its lattice periodic constant. Therefore, we have obtained a deep-subwavelength band gap by using Helmholtz resonant air cavities. The Zak phase is introduced to define the topological characteristics of band structures. The Zak phase of each band is (28, 29):

$$\theta_{Zak} = i \int_{-\pi/a}^{\pi/a} \langle u_k | \partial_k | u_k \rangle dk \qquad (1)$$

where $a$ is the lattice period, $u_k$ is the periodic Bloch function and $\partial_k$ is the partial derivative with respect to wavevector $k$. The detailed calculation process of the Zak phase is given in Note 2 in *supporting information*. The calculation results show that the Zak phase $\theta$ of the first and second band in Figure 2(d) is 0. We reconstruct the meta-molecule to be an expanded one with $d=1.2a$, and the degenerated point will be also lifted, as shown in Figure 2(e). In this case, the Zak phase $\theta$ of the first and second band in Figure 2(e) is $\pi$. Phase diagram of the meta-molecule in terms of the distance $d$ from $0.6a$ to $1.4a$ is shown in Note 3 in *supporting information*. To explore the topological interface state, we calculate the eigenfrequencies of the ordinary and topological acoustic chains. The ordinary chain consists of shrunk meta-molecules. The topological one is a chain consists of both shrunk and expanded meta-molecules. The numerically evaluated eigenfrequencies are plotted in Figures 2(f) and 2(g). The orange regions represent the band gaps. The black points represent the bulk states. Compared with the ordinary chain, we can find that there is a new eigenstate marked by a red point in the band structure of the topological chain. This is the topological interfacial state with a value of 19.74 kHz.

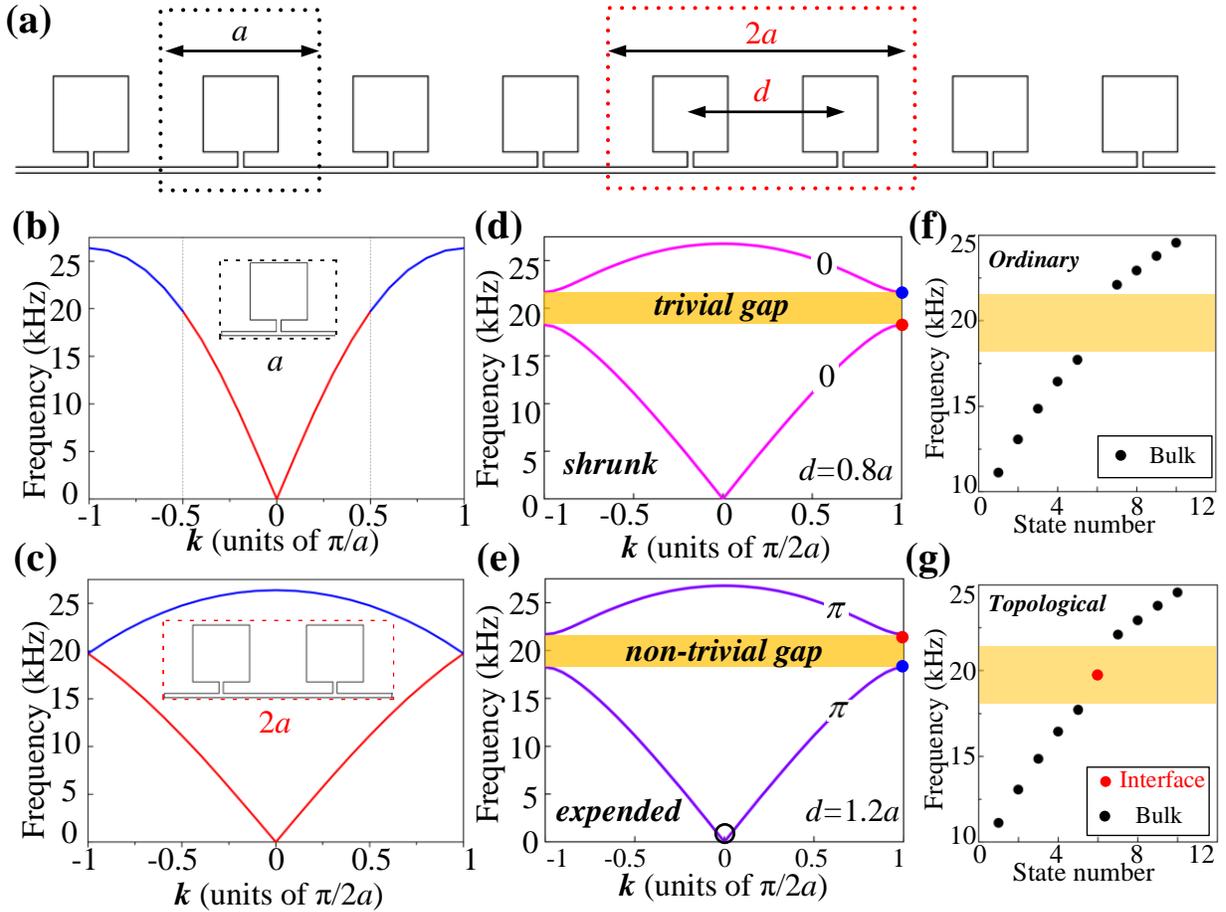

Figure 2. (a) Schematic of the acoustic metamaterials consisted of Helmholtz resonant air cavities. (b) Band structure for the previous unit cell with one Helmholtz resonant air cavity. (c) Band structure for the meta-molecule with two Helmholtz resonant air cavity. Band structures of meta-molecule with an interval $d=0.8a$ (d) and $d=1.2a$ (e). The Zak phase of each band has been labelled in the band structure. Numerically evaluated eigenfrequencies for ordinary (f) and topological (g) acoustic metamaterials. The red and black points represent the interfacial and bulk

states, respectively.

**Observation of interfacial topological state**

We mark the pressure of the air cavity is $p$, and the area of the water-air surface is $S$. Therefore, the oscillation force $F$ is proportional to the product of $p$ and S ($F \propto p \cdot S$). The surface tension coefficient of the water is $\sigma$. Therefore, the displacement amplitude of the oscillating water-air surface $\varepsilon$ will follow the positive relation $\varepsilon \propto (p \cdot S / \sigma)$. The formula indicates that the sound pressure can be measured indirectly by observing the displacement of the water-air surface. As shown in Figure 3(a), we use the high-speed camera to measure the micro-oscillation of the water-air surface (video 3, *supporting information*). The vibration excitation is incident from the left side of the microfluidic chip, which is driven by a piezoelectric transducer. We measure the displacements of the interfacial (position $L_1$, Figure S2 in *supporting information*) and lateral (position $L_5$) water-air surfaces, and the experimental results are shown in Figure 3(b). The blue and magenta curves represent the transmission spectrum of interfacial and lateral water-air surfaces, respectively. The interval of sampling frequency is 100 Hz, and the input voltage is 5 Vpp in the experiments. The magenta curve exhibits a low transmission efficiency in the frequency region [19.5 kHz, 22 kHz], which corresponds to the band gap of the ordinary chain. We note that there is a peak value of the blue curve, which is induced by the topological interfacial state in Figure 3(b). The frequency of this experimentally tested transmission peak is 20.0 kHz. We also experimentally measure the vibration amplitudes of the water-air surfaces near the topological interface under 20.0 kHz, as shown in Figure 3(c). We can find that the displacement at the interface is much larger than those at the lateral parts. It indicates that we are able to control the amplitude distribution of the water-air surface through acoustic topological insulator. We discuss the robustness of the topological state in Figure S4 and Table 1 (Note 4, *supporting information*).

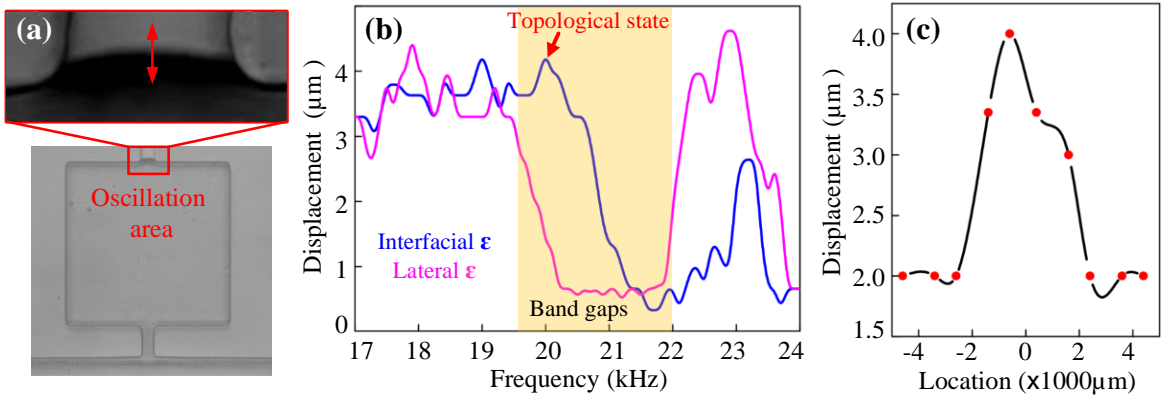

Figure 3. (a) Experimental observation of the oscillating water-air surface. (b) The experimental transmission spectrum of the interfacial and lateral water-air surface oscillating displacement. (c) The experimental displacement of the water-air surface when the incident frequency is 20.0 kHz, and the input voltage is 5 Vpp.

When the periodic Helmholtz resonant air cavities are replaced by the sidewall microcavities with width 40 μm, the theoretical natural frequency of microbubbles formed in microcavities is 167.5 kHz (Note 5, *supporting information*). Meanwhile, when the width of sidewall microcavities is larger than 100 μm, it will dramatically increase the difficulty to form microbubbles. Therefore, for sidewall microbubbles, it is almost impossible to obtain a natural frequency below 60kHz (Note 5, *supporting information*). However, we get a theoretical natural resonant frequency at 19.74 kHz by using the periodic Helmholtz resonant air cavities in our tweezers. It indicates that we can deeply decrease the resonant frequency of oscillating water-air surface, providing a technique to manipulate the large microparticles. Furthermore, the experimentally tested resonant frequency is 20.0 kHz which well matches with the numerical simulation result. It shows that the topological states of our tweezers are robust against manufacturing errors and environmental disturbances.

**Microparticle manipulations by the oscillating water-air surface**

We analyze the microparticle motions induced by oscillating water-air surfaces which can generate steady micro-vortices. Considering the density difference between the fluid (water, 1000 kg/m$^3$) and microparticles (polystyrene, 1050 kg/m$^3$), we can safely ignore the interaction of gravity and buoyancy. When the microparticles are near the oscillating water-air surface, they undergo both acoustic radiation forces and acoustic streaming forces. The acoustic radiation force is the time-averaged force steaming from the scattering of acoustic waves from the microparticle. The acoustic streaming force is induced by the viscosity between the moving microparticles and the streaming flow. The time-averaged acoustic radiation force exerted on the spherical microparticles can be estimated by (30):

$$\boldsymbol{F}_\mathrm{R} = \frac{4}{3}\pi\rho_\mathrm{f}\phi(\rho) \cdot \frac{R_a^{\,4} R_\mathrm{p}^3}{d^5} \omega^2 \varepsilon^2 \tag{2}$$

$$\phi(\rho) = 3(\rho_\mathrm{p} - \rho_\mathrm{f}) / (2\rho_\mathrm{p} + \rho_\mathrm{f}) \tag{3}$$

where $R_a$ is the effective radius of the air-cavity, $R_\mathrm{p}$ is the radius of the microparticles, $d$ is the distance between the centers of the air-cavity and the microparticle, $\omega$ is the driving frequency of the acoustic wave, $\varepsilon$ is the oscillating amplitude of the air cavity interface. $\phi(\rho)$ is the density contrast factor, and $\rho_f$ and $\rho_p$ represent the density of the surrounding fluid and the microparticle, respectively. Based on the sign of the density contrast factor, the acoustic radiation force can be either positive or negative. When the density contrast factor $\phi(\rho) > 0$, namely the density of microparticle is larger than that of the fluid, the high-density particle suffers the attractive acoustic radiation force and moves towards to the water-air surface. When the density contrast factor $\phi(\rho) < 0$, namely the density of microparticle is smaller than that of the fluid, the low-density particle suffers the repulsive acoustic radiation force and moves away from the water-air surface. In this paper, the polystyrene microparticle's density is 1050 kg/m³, 5% higher than that of the fluid. Therefore, the polystyrene microparticles will move towards to the water-air surface in our experiment (video 5, *supporting information*).

To further clarify the microparticle trapping mechanism, we discuss the acoustic streaming force which is induced by the fluid viscosity. The acoustic streaming force can be given by (30):

$$\boldsymbol{F}_{As} = 6\pi\mu R_p \boldsymbol{u} \quad (4)$$

Where $\mu$ (water, $1.01 \times 10^{-3}$ Pa·s) and $\boldsymbol{u}$ are the fluid dynamic viscosity and the velocity of the microparticle relative to the surrounding fluid, respectively. There is a ratio estimating the relative amplitude between the acoustic radiation force and the acoustic streaming force (30):

$$\boldsymbol{F}_R / \boldsymbol{F}_{As} \approx \rho_f \mu^{-1} \phi(\rho) R_p^2 \omega \quad (5)$$

For a resonant frequency 100 kHz which can be produced by microbubbles, we can obtain that the acoustic radiation force $\boldsymbol{F}_R$ equates to the acoustic streaming force $\boldsymbol{F}_{As}$ when microparticle size is 11.5 μm, seeing Note 6 in *supporting information*. This indicates that when the particle size is bigger than 11.5 μm, the acoustic radiation force will play the predominant role and microparticles (sizes bigger than 11.5 μm) are trapped on the oscillating water-air surface. Thus, it is difficult for microparticles with sizes larger than 11.5 μm to rotate following the streamline force. However, the rotational motion, especially for the coherent angular motion, plays a critical role in the normal development and the malignant transformation of cells(31). Thus, it is extremely meaningful to develop new acoustic technology to rotate these cells whose diameters mainly range from 10 μm to 30 μm. We can find that the threshold value of microparticles is 25.8 μm when the resonant frequency decreases to 20 kHz (Note 6 in *supporting information*). This indicates that the acoustic streaming force plays the predominant role compared with the acoustic radiation force, when the size of microparticle is smaller than 25.8 μm. Namely, we provide a new platform to rotate the large microparticles or cells whose sizes are up to 25 µm.

For the microparticle with a diameter at 20 µm, we can find that the acoustic streaming force is larger than the radiation force $(F_R / F_{As} \approx 0.6)$, which indicates that the acoustic streaming force plays a predominant role. Therefore, the 20 µm microparticles in our experiment will mainly follow the motion of acoustic streaming. The acoustic pressure distribution in microfluidics, induced by the oscillating water-air surface, is shown in Figure 4(a). We can observe that the acoustic energy gathers at the water-air surface, and the acoustic pressure intensity gradually decreases away from the surface. Figure 4(b) plots the distribution of the acoustic streaming velocity amplitude, and the red arrows represent the velocity direction. We can note that the oscillation water-air surface generates two vortices with opposite directions. We experimentally observe that a 20 µm particle does orbit rotation motion in the vortex, as shown in Figures 4(c)-4(h) (video 6, *supporting information*). The capture interval time is 1 ms, and the motion period is 40 ms. The corresponding rotation angles are respectively $90° \rightarrow 135° \rightarrow 220° \rightarrow 245° \rightarrow 315° \rightarrow 90°$ for Figures $(c) \rightarrow (d) \rightarrow (e) \rightarrow (f) \rightarrow (g) \rightarrow (h)$. We can find the velocity of microparticle's rotation near the water-air surface is large and the velocity away from the water-air surface is small, as the acoustic streaming force gradually decrease away from the water-air surface. We also achieve multiple microparticles rotation manipulation in Figure 5 (video 7 in *supporting information*).

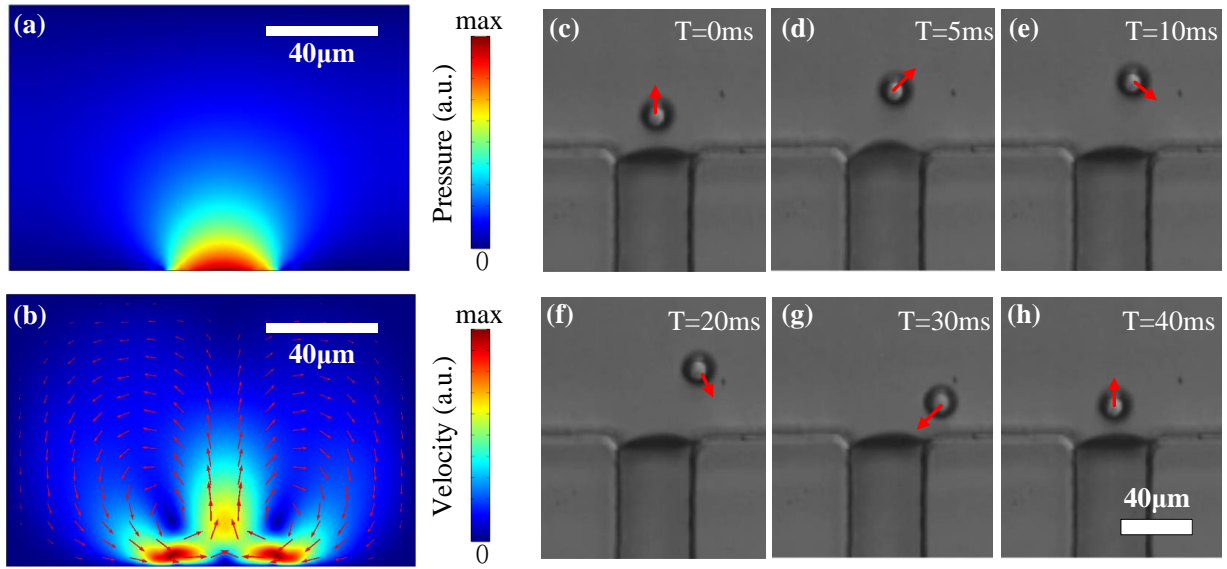

Figure 4. (a) Acoustic pressure distribution induced by the oscillating water-air surface. (b) Velocity of the acoustic streaming. (c)-(h) The trajectory distribution of 20 µm microparticle at different times. The driven frequency is 20.0

kHz, and the input voltage is 20 Vpp.

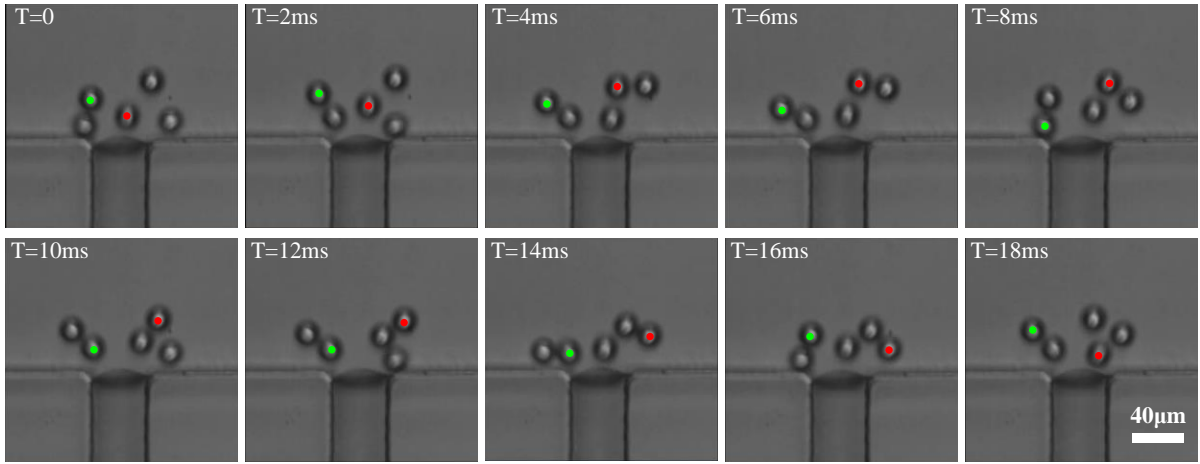

Figure 5. The trajectory distribution of multiple 20 μm microparticle at different times. Color points assist in tracking for observation. The driven frequency is 20.0 kHz, and the input voltage is 20 Vpp.

**Conclusion**

In this paper, we proposed a topological on-chip acoustic tweezer based on the topological interface states and experimentally observed great oscillations of water-air surfaces induced by acoustic waves in a microscale level. As the interface state is topologically protected, our tweezer is robust against manufacturing errors and environmental perturbations. Compared with microbubbles, the Helmholtz resonant air cavities in our tweezers greatly reduce the resonant frequency of the oscillating water-air surface, providing the possibility to drive large-scale microparticles or cells to do orbit rotation near the oscillating water-air surfaces. Thus, our topological on-chip acoustic tweezers provide a robust and subwavelength platform for microparticle manipulation on-chips, which have enormous potential applications in biomedical areas, such as contactless cell trapping, tissue culture, accurate drug release and targeted therapy.

**Acknowledgments**

This study was supported by the National Natural Science Foundation of China (Grants No. 11572121 and No. 51621004).